# High-output CPP-GMR sensor with synthetic-ferrimagnet free layer and enhanced spin-torque critical currents

M.J. Carey, N. Smith, S. Maat and J.R. Childress.

San Jose Research Center, Hitachi Global Storage Technologies, 3403 Yerba Buena Rd, San Jose, CA 95135

**Abstract:**

It is shown that the maximum stable output of a CPP-GMR sensor is increased significantly by using a synthetic ferrimagnet free layer, provided the electron current flows from free layer to reference layer. This free layer allows a larger magnetoresistance ratio for a given free layer magnetic moment, and in addition results in a greater than three-fold increase in the critical current above which spin-torque instability of the free layer occurs. In read heads with net free layer moments equivalent to only 4.5nm of $Ni_{80}Fe_{20}$, this effect is shown to result in sustainable sense current densities above $2 \times 10^8$ A/cm$^2$.

Spin-valves (SV) have been used in magnetic recording since 1998, when the current-in-plane (CIP) giant magnetoresistance (GMR) SV sensor was introduced [1]. More recently the current-perpendicular to the plane (CPP) tunneling magnetoresistance (TMR) sensor[2] has become standard. However, in the CPP geometry the sensor resistance increases with the smaller sizes required for increasing areal densities. Today's sensor dimensions are already below 60nm for >300 Gb/in$^2$ recording which has prompted a major effort to reduce the large resistance-area (RA) product of TMR sensors to below 1 $\Omega$-$\mu m^2$. Even so, the high impedence of TMR sensors at foreseeably smaller dimensions, and the accompanying excessive noise and degraded high-frequency performance, motivates a return to a metallic-GMR SV sensor in the CPP geometry (RA < 0.1 $\Omega$-$\mu m^2$). However at these low RA values, substantial sense current densities (>1x10$^8$ A/cm$^2$) are required to generate sufficient output voltages, and the interaction of the conduction electron spin with the sense and reference layer magnetizations can result in spin-torque (ST)-induced instability[3] which can render a CPP-GMR SV sensor nonfunctional. Therefore methods of retarding the onset of such instability is a key goal for the application of CPP-GMR spin-valves to recording sensors. In this Letter, we show that the critical sense current for ST-induced instability of the free layer can be dramatically increased by the use of a synthetic-ferrimagnet free layer and the correct choice of current direction. This approach is shown to enable CPP-GMR read heads with sense current densities up to 2x10$^8$ A/cm$^2$.

Several methods have already been demonstrated to reduce spin-torque effects in CPP-GMR sensors. One method is to use a "dual-SV" structure with a symmetric arrangement of two reference layers, which cancels the net spin-torque on the central free layer[4,5]. However the thicker sensor stacks of a dual-SV limits the linear resolution of the sensor, problematic for reaching high densities. Another method is to increase the magnetic damping of the free layer, either by rare-earth dopants[6] or cap layers[7], or by the spin-pumping effect using cap layers such as Pt[8]. Spin-torque effects can also be reduced by increasing free layer saturation magnetization ($M_s$) and/or total magnetic moment, but these values are typically constrained by the desired magnetic and/or transport properties of the sensor within existing recording system and other design considerations.

A known alternative free layer structure, but whose behavior under spin-torque excitations is first studied here, is the synthetic-ferrimagnet (SF) free layer or antiparallel (AP) free layer[9]. In the SF-FL structure (Fig.1 inset) the free layer is a multilayer of the type FL1/APC/FL2, where APC is an antiparallel coupling layer such as Ru and FL1 and FL2 are two separate magnetic layers. This configuration is similar to the widely-used synthetic–antiferromagnet (SAF) pinned layer structure formed by pinned layer/APC/reference layer, except that the SF-FL structure has a nonzero net magnetic moment $m_{FL}=m1-m2$, where $m1$ and $m2$ are the magnetic moments/area of FL1 and FL2, respectively. This allows, for a given $m_{FL}$, to use a relatively thick FL1 thickness, which can increase the CPP magnetoresistance of the spin-valve[10]. In addition, it is of interest to evaluate the influence of the additional FL2 layer on the overall ST-induced excitations of the free layer.

Spin-valve films for the present work were deposited by magnetron sputtering onto Si substrates as discussed elsewhere[4]. For magneto-transport measurements, multilayer films were patterned into pillars with nominal diameters between 300 and 50nm using e-beam lithography and Ar$^+$ ion milling. The full structure for the devices described here was underlayer/5 Ta/1.5 Cu/7 IrMn/3 CoFe/0.55 Ru/1 CoFe/0.4 Cu/1 CoFe/0.4 Cu/1 CoFe/5 Cu/FL or SF-FL/1 Ru/cap layer, with all thicknesses in nm. The CoFe composition is $Co_{50}Fe_{50}$ (atomic %). The SF-FL structure was 0.6 CoFe/4+t NiFe/0.2 CoFe/0.55 Ru/0.2 CoFe/t NiFe, resulting in a constant net $m_{FL}$ of 0.36memu/cm$^2$, equivalent to 4.5nm of $Ni_{80}Fe_{20}$ ($M_s$=800 emu/cc). A control structure with a single free magnetic layer (FL) of 0.6 CoFe/3.8 NiFe/0.2 CoFe (also with a $m_{FL}$=0.36memu/cm$^2$,) was used for comparison. Magnetoresistance measurements were performed quasistatically in ±1.5kOe applied fields using a constant voltage, where a positive voltage is defined to produce an electron current as shown in Fig.1. $\Delta RA$ values were derived from the resulting $R$-$H$ curves, where $R_{min}$ is the resistance of the device when the reference layer and FL1 magntizations are parallel (P-state), $\Delta R$ is the change in resistance upon switching FL1 to be antiparallel with the reference layer (AP-state), and $A$ is the device area. $R_{min}$ vs. 1/$A$ plots were used to correct the resistance values for lead resistances and lithographic windage effects. The $RA_{min}$ value of all the structures is about 40 mΩ-μm$^2$.

Fig.1 shows the value of the magnetoresistive signal $\Delta RA$ (measured at -10mV) as a function of the FL2 NiFe thickness t. $\Delta RA$ is notably higher for the AP-free layer structures as compared to the control structure, increasing from 0.68±0.04 to 0.93+0.03 mΩ−μm$^2$ for devices with a FL2 NiFe thickness t > 1.5nm. This corresponds to $\Delta R/R_{min}$ values increasing from 1.7 to 2.7%. For CPP-GMR devices, it might be expected that the FL2 layer will contribute a *negative* value to the total $\Delta RA$ due to its antiparallel magnetization orientation compared to FL1. However this is not noticeable here, consistent with the short spin-diffusion length ~4nm previously estimated for NiFe[11] as the NiFe thickness of FL1 is >4.5nm for all SF-FL samples (FL1 also includes two CoFe nanolayers).

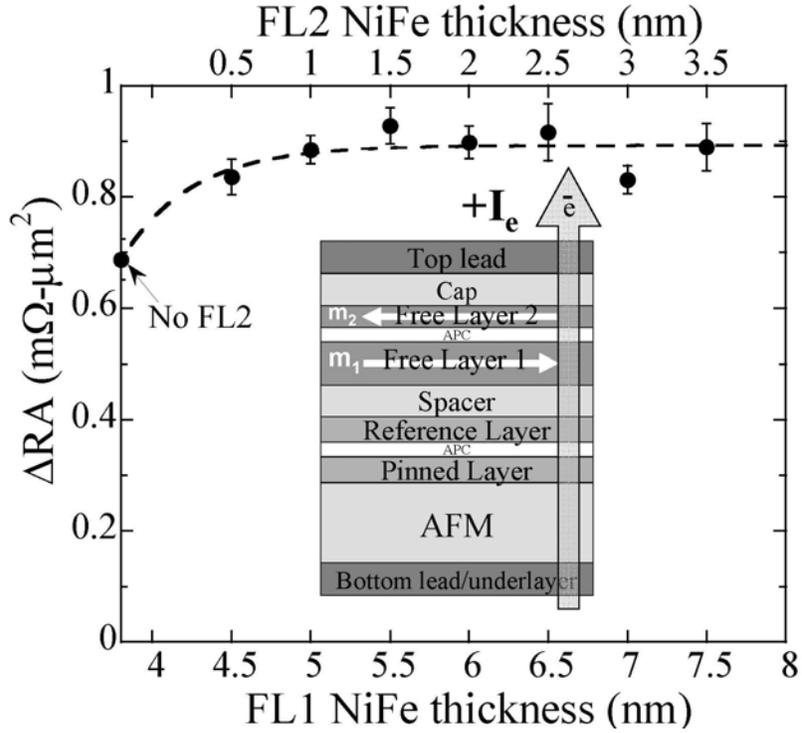

Fig.1: ΔRA vs. NiFe thickness in FL1 (bottom axis) and FL2 (top axis). Inset: Spin-valve structure with synthetic-ferrimagnet free layer. For transport measurements, positive electron current is defined as electron traveling from reference layer to free layer.

Spin-torque stability was examined using lock-in measurement of $R' = dV/dI_e$ vs. $I_e$ (tickle-current = 40 μA) in "hexagonal" device with cross sectional area similar to that of a 75nm diameter circle. Two examples of these measurements, along with several corresponding $\Delta R$-$H$ loops (with $H$ collinear to the long hexagon axis), are shown in Fig. 2. At sufficiently small $H$ and/or $I_e$, these devices are bi-stable due to the uniaxial shape anisotropy and both $\delta R$-$H$ (at low $I_e$) and $\delta R'$-$I_e$ loops (at $H = 0$) show the expected hysteresis. The latter, in particular, indicate that the *overall ST-stability of the FL, when employing positive (negative) sense current, is then limited by its positive (negative) critical current $I_e^{\text{crit}}$ in the AP-state (P-state).*

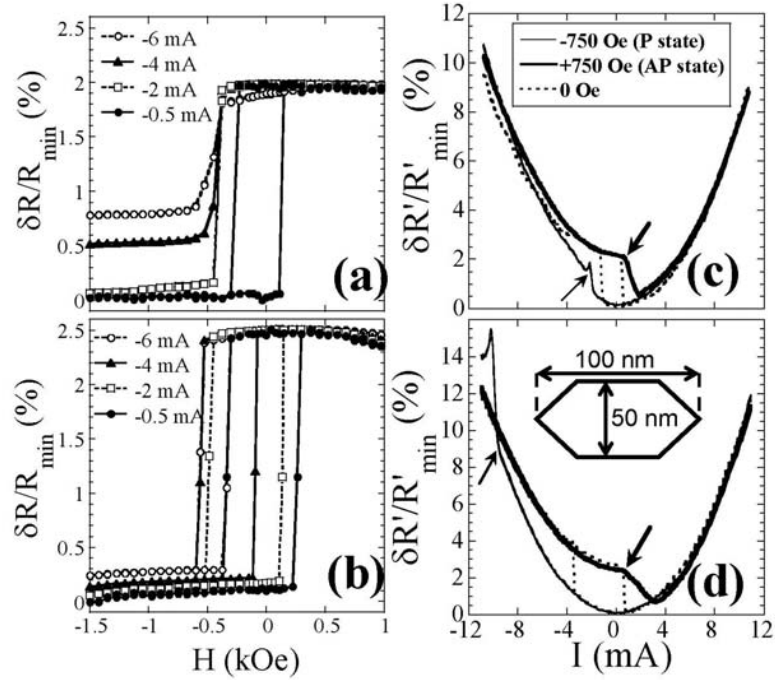

Fig.2: $\delta R - H$ transfer curves at discrete $I_e$ as indicated (left), and $\delta R' - I_e$ transfer curves at $H = 0, \pm 750$ Oe (right), where $R' \equiv dV/dI_e$. The value of $R_{min}$ or $R'_{min}$ is subtracted out, and the $\delta R - H$ curves are further aligned at $\delta R_{max} \to \Delta R$ to remove thermal shift. $t_{FL2} = 0.5$ nm NiFe for (a) and (c), $t_{FL2} = 2.5$ nm NiFe for (b) and (d) The arrows in (c) and (d) indicate the onset of spin-torque instability for $H = \pm 750$ Oe

However, $I_e^{crit}(H=0)$ can depend on the device coercivity $H_c(I_e \to 0)$ (which itself is dependent on device geometry that is irregular at 50 nm dimensions), and is susceptible to thermal[12] and self-field effects. A more reliable measurement of critical currents $I_e^{crit}$ can be achieved using external fields $H \gg H_c(I_e \to 0)$, such as shown in Fig. 2c,d with $H \cong \pm 750$ Oe, where *at most* only one state (P or AP) is both magnetostatically and ST-stable. Here, The values of $I_e^{crit}$, denoting the start (from $I_e = 0$) of continuous instability, are determined by observing the onset of significant deviation from the otherwise parabolic shape of $\delta R'(I_e)$ due to Joule heating. These deviations, discernable by inspection for the (nonhysteretic) $\delta R'(I_e)$ shown in Fig. 2, are symptomatic of continuous, precessional-like motion of the FL magnetization. It is generally accompanied by a rapid increase in low-frequency $1/f$-like noise which serves as an alternative (usually more sensitive) detection technique as was shown earlier[13] and employed recently.

Fig. 3 shows a summary of all measured $I_e^{crit}$ vs FL2 thickness, for devices initially stable in either P or AP states. The enhancement in *negative* $I_e^{crit}$ (in P-state) with increasing FL2 thickness is *dramatic*, and dramatically *larger* than that observed for positive $I_e^{crit}$ (in AP-state). These results are highly consistent (both qualitatively and quantitatively) with further measurements on 75-nm diameter circular devices of very similar cross sectional area. In addition, the enhancement with FL2 thickness of negative values of $I_e^{crit}$ for the unidirectionally P-stable devices with $H = -750\,\text{Oe}$ is also seen to be substantially larger than that observed when uniaxially bi-stable at $H = 0$. A physical explanation, along with a simple macrospin model which reproduces all of these observations, is described elsewhere[14].

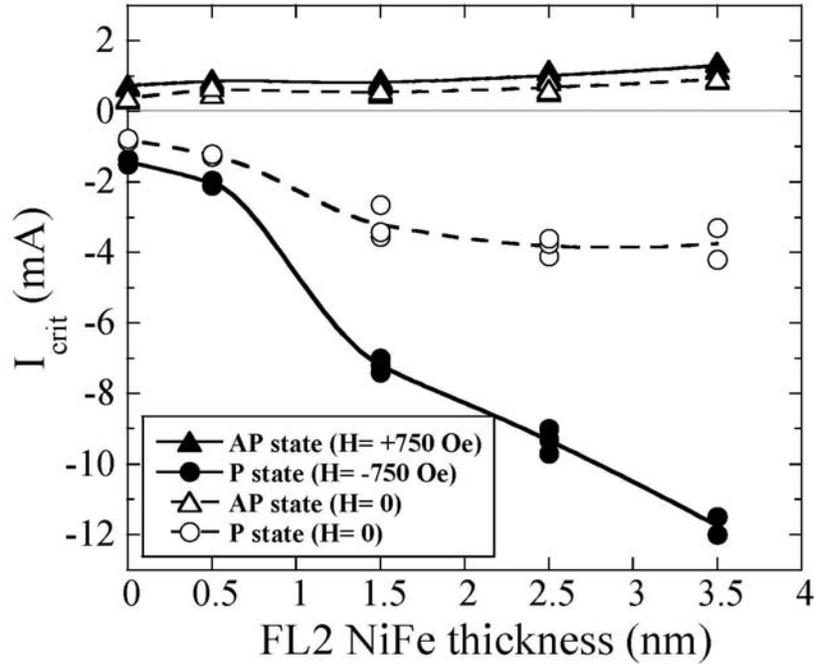

Fig.3: Spin-torque critical current as a function of FL2 thickness *t* for spin-valve with synthetic-ferrimagnet free layer structure and constant net magnetic moment. Two or three devices are tested for each value of t, with *t*=0 corresponding to a single-FL control structure. The mean resistance of these devices is about 10Ω.

As a practical matter, one is free to choose the polarity of the sense current, for which negative $I_e$ is clearly preferable for the single magnetic layer FL design. To test device performance under recording test conditions, films similar to those described above were fabricated into stabilized read-only heads using a combination of ebeam and optical lithography down to ~ 30nm track-width sizes. For this experiment the FL2 NiFe thickness t was chosen to be 2.5 nm, so that the corresponding FL1 NiFe thickness was 6.5 nm. This resulted in device-level $\Delta R/R_{min}$ values of about 2.5%. Fabrication details and test conditions were similar to those previously described. Under these conditions, the behavior of the sensor as a function of voltage bias was tested to evaluate the sensor behavior with respect to ST-induced excitations. Fig. 4 shows the output amplitude vs. bias for various heads with ~30-40nm physical track-widths and an RA value of about 40 m$\Omega$-$\mu m^2$. For positive bias, transfer curves become rapidly distorted with voltages as low as 15-20 mV, as evidenced by the decreased amplitude with increasing bias. For negative bias however, the amplitude continues to increase with bias up to values in excess of 100 mV. This corresponds to current densities of about $J_{max}$ ~ 2.5x10$^8$ A/cm$^2$, which is more than twice the maximum values we have previously reported for CPP dual-spin valves, which, in turn, are higher than those observed in single-spin valves with non synthetic-ferrimagnet free layers. These high values of $J_{max}$ illustrate the effectiveness of the SF-FL structure in suppressing the effect of spin-torque on the free layer for CPP-GMR sensors.

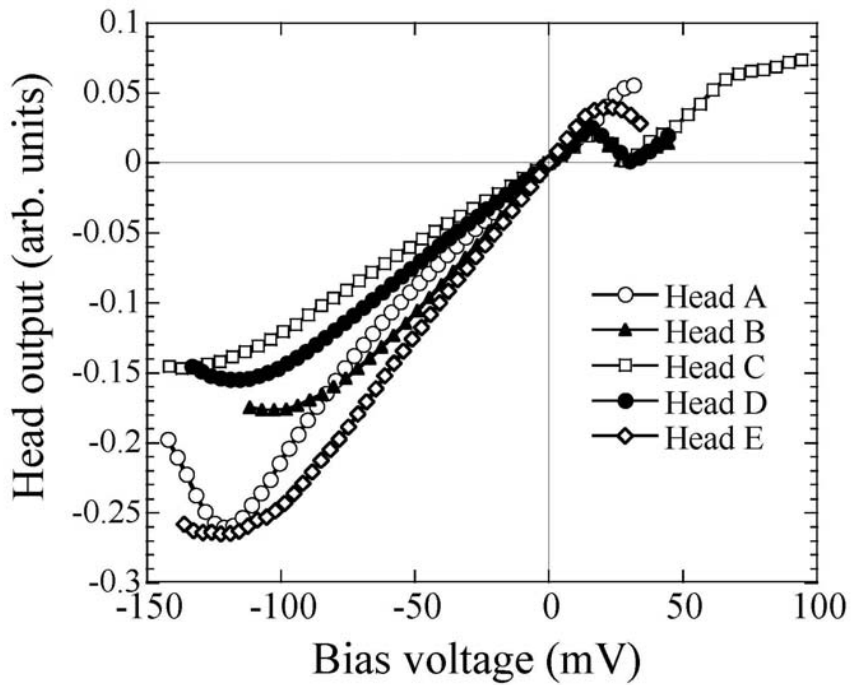

Fig.4: Read-only recording head output amplitude vs. bias voltage applied to the sensor. These heads include a synthetic-FL structure similar to that described here with FL2 NiFe

thickness ~ 2.5nm.  Only with negative bias can substantial current densities be achieved before the onset of excessive ST noise.

In conclusion, we have shown that the use of a synthetic ferrimagnet free layer structure for CPP-GMR sensors, together with a sensor current where electrons flow from the free layer to the reference layer, results in a substantial increase in the critical current density $J_{max}$ to greater than $2\times10^8$ A/cm$^2$.  Along with increased spin-valve signal ΔRA, increasing $J_{max}$ in heads results in an increase in sensor output voltage $\Delta V = \Delta RA \times J_{max}$, a key requirement for improving the applicability of CPP-GMR to high-density magnetic recording sensors.  Consequently, synthetic–ferrimagnetic free layers in CPP-GMR sensors constitute a promising route for the next generation magnetic recording sensors at 500 Gb/in$^2$ and beyond.

The authors thank Ching Tsang and Jim Moore for valuable assistance with recording head characterization.